\documentstyle[epsfig,aps,prl,twocolumn]{revtex}
\begin{document}

\wideabs{

\title{Phonon-assisted electronic topological transition in MgB$_2$ under pressure}

\author{ Viktor V. Struzhkin, Alexander F. Goncharov, Russell J. Hemley, Ho-kwang Mao}

\address{Geophysical Laboratory and Center for High Pressure
Research, Carnegie Institution of Washington, \\
5251 Broad Branch Road NW, Washington D.C. 20015 U.S.A}

\author{G. Lapertot \cite{a1}, S. L. Bud'ko, and P. C. Canfield}

\address{Ames Laboratory and Department of Physics and Astronomy,
Iowa State University, Ames, IA 50010}


\maketitle

\begin{abstract}

We report measurements of the superconducting critical temperature $T_c$  
of polycrystalline MgB$_2$ samples containing isotopically pure $^{10}$B and  $^{11}$B
under quasi-hydrostatic pressure conditions in He pressure media up to 44 GPa.
Measurements to volume compressions  $V/V_0 \sim$  0.82 allow us 
to observe a kink in the volume dependence of  $T_c$  for Mg$^{10}$B$_2$ (at 20 GPa) and 
Mg$^{11}$B$_2$ (at 15 GPa). The pressure dependence  of the $E_{2g}$ mode also 
changes abruptly around 20 GPa for the  Mg$^{10}$B$_2$ sample.  The anharmonic character of the 
$E_{2g}$ phonon mode and anomalies in $T_c$ pressure dependence are interpreted  
as the result of a phonon-assisted  Lifshitz electronic topological transition.

\end{abstract}
}

The recently discovered high-temperature superconductor
MgB$_2$ has attracted considerable interest in condensed-matter and 
materials physics \cite{Akimitsu}. 
Experiment \cite{bud'ko,hinks} as well as theory \cite{kortus,an,kong,yildirim} 
indicate that MgB$_2$ can be treated as a phonon
mediated superconductor. Calculations show that 
the strongest coupling is realized for the phonon branch in the Brillouin zone 
from $\Gamma$ point ($E_{2g}$ phonon) to $A$ point ($E_{2u}$ phonon), 
which is  related to vibrations of the B atoms \cite{an,kong,yildirim}.
This makes MgB$_2$ a unique system with a single phonon branch dominating 
the superconducting properties within the framework of phonon-mediated mechanism 
for superconductivity. By knowing the pressure dependence of these phonons frequencies
and pressure dependence of $T_c$, the electron-phonon coupling in this material 
can be directly addressed. 

The pressure 
dependence of the $E_{2g}$ phonon was measured recently in our laboratory 
using Ne pressure media \cite{goncharov}. 
The phonon mode is very anharmonic: it has large Raman linewidth 
and an unusually high Gr\"uneisen parameter. Theoretically, the anharmonicity 
of this mode  has been understood as arising from
its strong coupling to the partially occupied planar B $\sigma$ bands
near the Fermi surface \cite{yildirim}. 
The theoretical calculations presented in Ref.\cite{kunc} predict 
high Gr\"uneisen  parameter values for the whole phonon branch from $E_{2g}$   
to $E_{2u}$ phonon, and the frequency for the $E_{2g}$ phonon with anharmonic 
corrections was predicted to be close to 570 cm$^{-1}$.
The  effects of pressure on superconductivity in MgB$_2$ studied to 1.84 GPa
in fluorinet \cite{lorenz} and 0.7 GPa in a He pressure medium \cite{tomita} show a
decrease of $T_c$  with the rate of 1.6 K/GPa and 1.11 K/GPa,
respectively.  Other measurements \cite{lorenz2} 
in He pressure media showed pressure coefficients 1.45 K/GPa, and 1.08 K/GPa 
for different samples.
The pressure studies in a Bridgman anvil apparatus with steatite 
as a pressure medium showed  different results for sintered and  powder 
samples, with a highly nonlinear pressure dependence measured for sintered sample   
\cite{monteverde}). Magnetic susceptibility studies in diamond anvil cells with  
alcohol pressure medium  have been performed to 28 GPa,
and an anomaly in the $T_c$(P) dependence was observed around 10 GPa  \cite{Tissen}, which 
was interpreted as a signature of the electronic topological transition (ETT) 
\cite{Lifshitz}.
Another diamond cell magnetic susceptibility experiment with a He medium 
 to 20 GPa did not confirm the anomaly at 10 GPa,
and showed a sublinear volume dependence of  $T_c$ \cite{Deemyad}. 
Our experiment  on Mg$^{10}$B$_2$ to 15 GPa in a He pressure medium \cite{goncha}  
showed a linear volume dependence of $T_c$, different from Refs.\cite{Tissen,Deemyad}.   
 
Compressibility data have been obtained by neutron diffraction (to 0.62 GPa)
\cite{jorgensen} and synchrotron x-ray diffraction (to 6.15 GPa
\cite{prassides}, 8 GPa \cite{vogt}, and 11 GPa \cite{goncharov}).  
Based on theoretical calculations
of the electronic density of states at the Fermi level, which show a very
moderate decrease with pressure, the dominant contribution to the decrease in
$T_c$ under pressure has been proposed to be due to an increase in phonon
frequency \cite{loa}. 

Here we report quasihydrostatic measurements of $T_c$ in He pressure 
media in MgB$_2$ with isotopically pure $^{10}$B up to 44 GPa,
and with isotopically pure $^{11}$B up to 33 GPa. 
Our low pressure data agree with previous 
measurements in hydrostatic conditions, with $dT_c/dP$=1.1 K/GPa. At higher pressures, 
however, $T_c$ deviates from a linear dependence as a function of pressure, although 
the volume dependence is linear for both isotopes to 15 GPa. Moreover, we observe a kink in the 
volume dependence of $T_c$  at 20 GPa for Mg$^{10}$B$_2$, and at 15 GPa for  Mg$^{11}$B$_2$.
The kink at 20 GPa for  $^{10}$B isotope correlates well with a similar 
abrupt change  in the pressure dependence of the $E_{2g}$ Raman mode, 
which suggests that  the $E_{2g}$ 
phonon is involved in the observed behavior of the superconducting transition.
We propose that pressure-induced changes in both the  $E_{2g}$ phonon mode and $T_c$ result from 
a phonon-assisted electronic topological transition \cite{Lifshitz}. This finding 
further means that the zero point 
motion of B atoms should be taken into account in treatment of the electronic structure, 
as opposed to standard approaches  which neglect the quantum character of the nuclei.

Samples of MgB$_2$ were similar to those used in Ref.\cite{bud'ko,finnemore}, 
with isotopically pure $^{10}$B and $^{11}$B.  They were essentially 
in a powdered form consisting of aggregates of 30-50 $\mu$m in linear dimensions,
 which is ideal for high-pressure experiments.  Our experiments were done with 
Be-Cu  nonmagnetic diamond anvil cells using magnetic susceptibility 
techniques \cite{technique}.  We used diamond anvils 
with a flat tip 400 $\mu$m in diameter.  A flake of MgB$_2$  approximately 
40 $\mu$m in diameter and 10 $\mu m$ in thickness was loaded into a 
200-$\mu$m  diameter hole, prepared in a nonmagnetic NiCr(Al) gasket \cite{technique}. 
He has served as the pressure transmitting medium.  
Pressure was increased at  40 to 50 K and measured using the standard ruby 
fluorescence technique. Thus, we can specify the conditions for this experiment as 
qusihydrostatic, because pressure was applied when the  He medium was solidified.  
However, our results show very good agreement with low pressure hydrostatic data, 
so we believe 
that nonhydrostatic effects in our experiment are negligible. We warmed the Mg$^{10}$B$_2$ 
sample 
to room temperature twice during this experiment (at 10 GPa and  25 GPa), 
and kept it at room temperature for few days.
This did not have any effect on the observed pressure dependence of $T_c$; we 
conclude that there is no substantial effect of thermal cycling on $T_c$.
Experiments with  Mg$^{11}$B$_2$ were done in a similar manner, with the only 
difference being that we warmed the sample to room temperature when the pressure was 
below 11 GPa to achieve hydrostatic conditions on compression. Above 11 GPa  the Mg$^{11}$B$_2$ 
sample was always kept below room temperature; however, it was warmed up occasionally 
to 120-130 K.
We also performed a nonhydrostatic experiment (without pressure medium) to 25 GPa 
with Mg$^{10}$B$_2$;
the observed pressure dependence differed from the run in the He  
medium, and resembled the data for powdered samples obtained by Monteverde et al 
\cite{monteverde}.  

\begin{figure}
\centerline{\epsfig{file=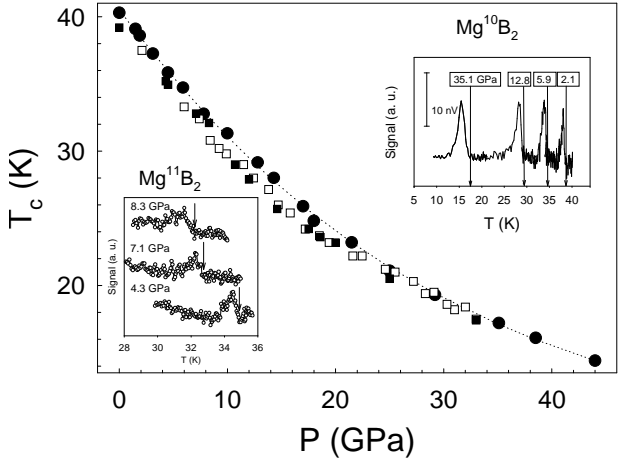,width=9cm}}
\caption{Pressure dependence of $T_c$ in a He medium. $T_c$ is identified 
as the onset of the magnetic signal \protect \cite{technique,technique2} as illustrated 
in the insets for selected pressures.   
The solid circles are data for \protect Mg$^{10}$B$_2$ (dotted line is guide to the eye). 
The data for  \protect Mg$^{11}$B$_2$  are shown as solid squares for compression, 
and open squares for decompression. No decompression data were obtained for the  
\protect Mg$^{10}$B$_2$ sample.}
\label{fig1}
\end{figure}

In Fig.\ref{fig1} we show the $T_c$ as a function of pressure; temperature 
scans at selected pressures are also shown. The signal observed is close to the 
limit of the sensitivity of our setup, which we have recently improved\cite{technique2}. 
The signal is superimposed on the 
nonlinear paramagnetic background from the gasket material at lower temperatures 
(below 25 K), which has a characteristic  ${{1}\over{T}}$ dependence (subtracted from the 
data) \cite{technique2}. However, the onset of $T_c$ can be reliably identified with  
an accuracy  0.2-0.8 K (depending on the actual quality of the data, 
as illustrated in Fig.\ref{fig1}) up to the highest pressures reached in this experiment. 

\begin{figure}
\centerline{\epsfig{file=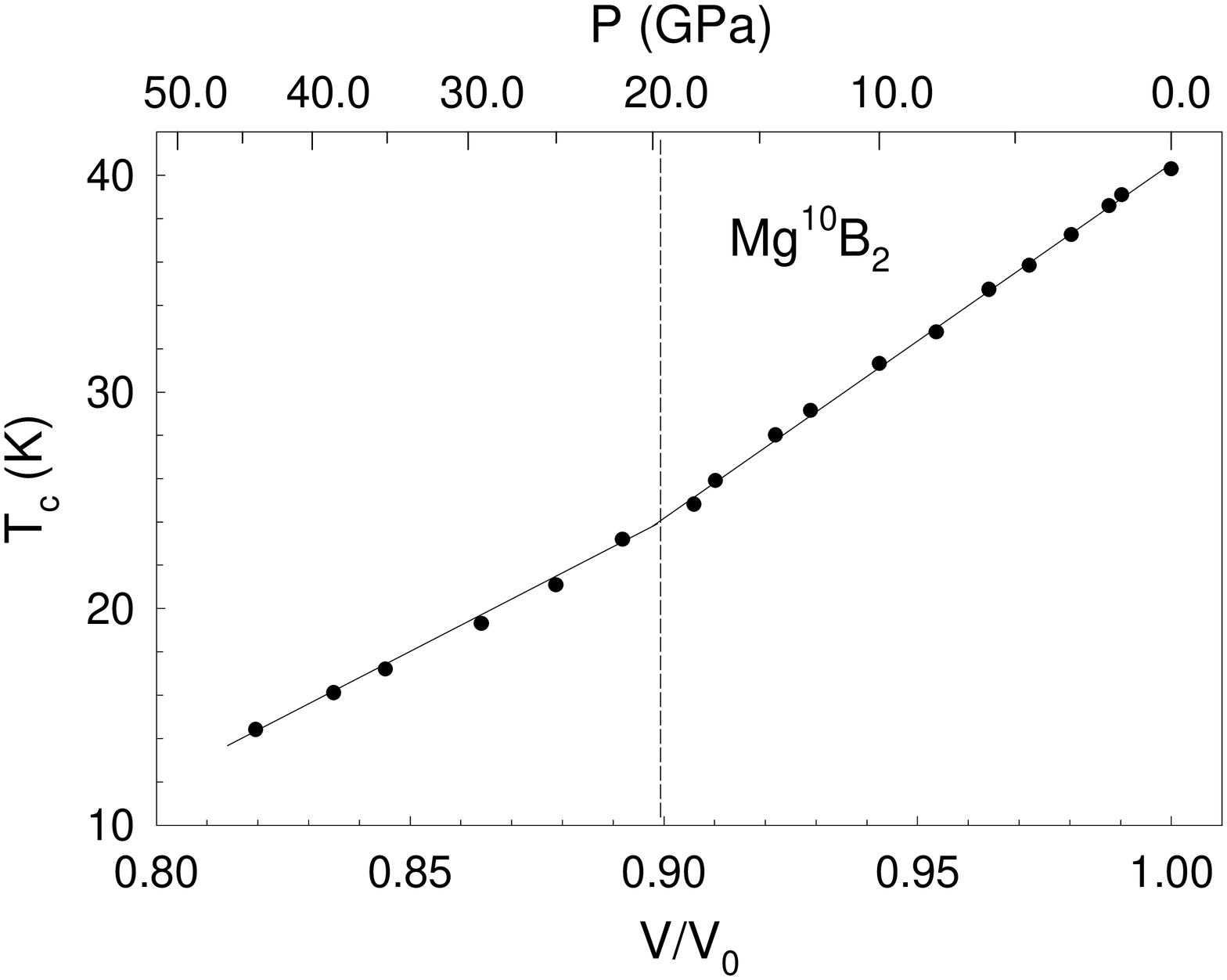,width=9cm}}
\caption{Volume dependence of $T_c$ for  \protect Mg$^{10}$B$_2$. 
The Vinet equation of state \protect \cite{vinet} 
was used to calculate the P-V relation with a bulk modulus B$_0$=150 GPa, 
and its derivative  B$_0$'=4.0. }
\label{fig2}
\end{figure}

\begin{figure}
\centerline{\epsfig{file=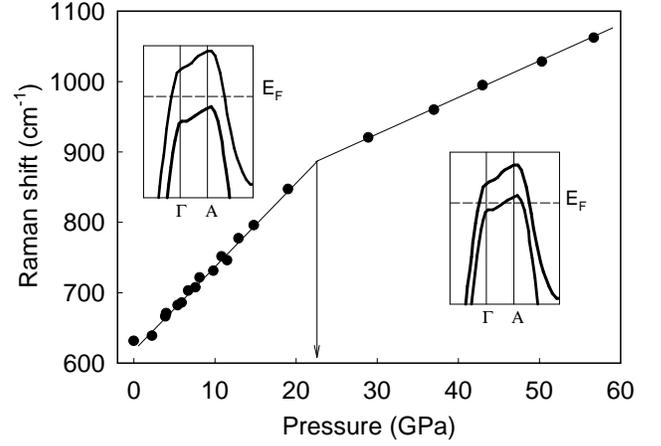,width=9cm}}
\caption{Pressure dependence of the \protect $E_{2g}$ Raman mode in \protect Mg$^{10}$B$_2$. The linewidth 
of the mode was  anomalously large around 20 GPa \protect \cite{goncha}.}
\label{fig3}
\end{figure}

We plot $T_c$ for Mg$^{10}$B$_2$ as a function of volume in Fig.\ref{fig2}.
One can clearly distinguish a kink in $T_c$(V) curve at a volume that corresponds 
to 20 GPa. We have measured  $E_{2g}$ Raman mode frequency at room temperature to 50 GPa 
to understand the anomaly in $T_c$, and we observed similar 
kink in pressure dependence of the Raman mode slightly above 20 GPa. 
The details of the Raman experiment will be published elsewhere \cite{alexraman}.
Pressure dependence of the  $E_{2g}$ Raman mode frequency is shown in Fig.\ref{fig3},
together with the cartoons which illustrate our understanding of this 
anomalous behavior of the superconducting T$_c$ and the E$_{2g}$ phonon mode. 

The  left inset in Fig.\ref{fig3} illustrates 
that at lower pressures the zero-point motion of B atoms for the $E_{2g}$ mode splits 
the boron in-plane $\sigma$ bands strongly, so that the lower band moves below 
the Fermi level, thereby crossing it and fulfilling the condition for ETT. 
This means that for a frozen-phonon calculation there should be an anomalous 
contribution to the  total 
energy, which behaves similar to suggested by Lifshitz \cite{Lifshitz} 
$2{{1}\over{2}}$ power  term in the 
free energy, with the amplitude of the phonon mode being a parameter, which drives 
the electronic subsystem through such a transition. As a result, the phonon frequency
will be strongly anharmonic, and its volume derivative (the Gr\"uneisen parameter) may 
even have an anomaly at the transition.          
At higher pressures the zero-point motion does not split  $\sigma$ band strong enough 
for the lower band to cross the Fermi level (right inset). Thus, system is always at  
conditions when there is no anomalous contribution to the free energy 
from ETT-like  2${{1}\over{2}}$ power terms, and the phonon mode and $T_c$ 
behave in a more regular manner.
Between those two regimes there should be a small pressure range in which the amplitude 
of the zero-point motion is just enough for the top of the lower band to coincide 
with the Fermi level. We propose that this condition is almost fullfiled at the 
observed kinks in pressure dependencies of $T_c$ 
and $E_{2g}$ phonon mode frequency.

\begin{figure}
\centerline{\epsfig{file=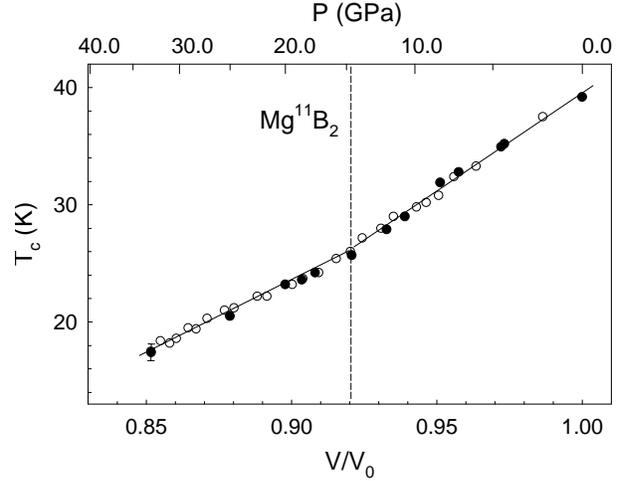,width=9cm}}
\caption{Volume dependence of $T_c$ in \protect Mg$^{11}$B$_2$. Full symbols are for compression; 
open symbols for decompression. We calculated the equation of state using the same parameters as 
for \protect Mg$^{10}$B$_2$. The variation of the bulk modulus by $\pm 5$~GPa does not shift volume 
position of the kink by more than $\delta$V/V$_0\sim$0.005.}
\label{fig4}
\end{figure}

The volume dependence of T$_c$ for Mg$^{11}$B$_2$  
is shown Fig.\ref{fig4}. A similar kink is clearly visible around 15 GPa. We believe that 
the lower pressure for the observed transition is due to the isotope effect: the 
zero-point motion for a heavier atom is smaller, and the matching condition 
for the  $\sigma$ band is fulfilled at lower compression of the lattice. 
We also  note that the transition at room temperature from Raman data 
for  Mg$^{10}$B$_2$  (Fig.\ref{fig3}) occurs at  slightly higher pressure, than the 
low-temperature transition as observed from the T$_c$(V) plot (Fig.\ref{fig2}). 
Although the difference is close to the observation error,  we may tentatively attribute 
this small effect to the lattice expansion: higher pressure is needed 
to compress the structure at room temperature to reach the transition. 
 
Several arguments support the observed isotope trend, 
following the reasoning proposed by An and Pickett \cite{an}. They noticed that 
the $\sigma$  bands belonging to boron, which form cylindrical Fermi surface 
sheets  \cite{kortus}, can be treated as quasi-two-dimensional. 
The states in these bands contribute most of the electron-phonon coupling responsible for 
superconductivity \cite{an}.
The overall splitting of the $\sigma$ band is characterized by $p-p$ matrix element  
$t_{pp\sigma} \sim d^{-3}$, where $d$ is B-B bond length. Thus, the deformation 
potential of the $\sigma$ band  will be proportional to the derivative of the above matrix 
element with respect to $d$ ($E_{2g}$ mode modulates B-B distance), 
and thus  $|\vec {\cal D}| \sim d^{-4}$.
We have determined earlier that the phonon frequency of the E$_{2g}$
mode scales as $\omega \sim (a/a_0)^{-10.8}= (d/d_0)^{-10.8}$ below 15 GPa  \cite{goncharov}. 
The amplitude of the zero-point motion $u \sim \omega^{-1/2}\sim (d/d_0)^{5.4}$, 
which means that the splitting of  $\sigma$  bands 
$\Delta E \sim \vec {\cal D} u \sim (d/d_0)^{1.4}$ is decreasing almost 
proportionally to the B-B bond length. 

The low-pressure regime for both isotope compounds suggests a strong contribution from 
ETT anomalies to the observed properties of the materials. It should be noted  
that the electron-phonon coupling may be strongly affected by the non-adiabatic 
effects due to the violation of the condition that the Debye frequency is much less than 
the Fermi energy  $\omega_D/E_F \ll 1$ \cite{Migdal} close to the ETT regime. 
We also expect large effects of uniaxial stresses and 
impurities on the pressure dependence of the superconducting transition in MgB$_2$.

In summary, we have measured the superconducting transition temperature in isotopically 
pure Mg$^{10}$B$_2$ and Mg$^{11}$B$_2$ under quasihydrostatic conditions 
(He pressure media) to 44 GPa. 
Although the initial linear slope of $T_c$ is found to be  in excellent agreement 
with previous hydrostatic experiments \cite{tomita,lorenz2}, 
we observed a  kink in the volume dependence of  $T_c$  at 20 GPa for Mg$^{10}$B$_2$, 
and at 15 GPa for  Mg$^{11}$B$_2$. A similar anomaly was found in the pressure 
dependence of the $E_{2g}$ phonon in  Mg$^{10}$B$_2$. We argue that observed 
anomalies are related to the phonon-assisted electronic topological transitions. 
This finding further means that the zero-point 
motion of the atoms should be taken into account in treatments of the electronic structure 
for this system. The nonhydrostatic stresses and impurity effects may be responsible 
for the most of the discrepancies in the available high pressure data for 
these intriguing materials.

We thank O. Gulseren and T. Yildirim  for the discussion of the phonon anharmonicity, 
and E. Gregoryanz for the experimental help. 
We also thank O. Gulseren for sharing his data on theoretical 
estimates of T$_c$(P) prior to publication.
We acknowledge financial support of CIW, NSLS, NSF and the Keck Foundation.
Work at Ames laboratory was supported by the Director for Energy Research,
Office of Basic Energy Sciences, U.S. Department of Energy.


\end{document}